\documentclass[conference]{IEEEtran}
\usepackage{cite}
\usepackage{amsmath,amssymb,amsfonts}
\usepackage{algorithmic}
\usepackage{graphicx}
\usepackage{amssymb,comment}
\usepackage{algorithmic}
\usepackage{algorithm}
\usepackage{caption} 
\usepackage{subfig}
\usepackage{subcaption}
\usepackage{scrextend}
\usepackage{verbatim}
\usepackage{color}
\usepackage{tablefootnote}
\usepackage{setspace}
\usepackage{lipsum}
\usepackage{geometry}
\usepackage{url}
\begin{document}
	
	\newtheorem{theorem}{\bf Theorem}
	\newtheorem{lemma}{\bf Lemma}
	\newtheorem{remark}{\bf Remark}
	\newtheorem{proposition}{\bf Proposition}
	\newcommand{\bs}[1]{\boldsymbol{#1}}
	\newcommand{\mbf}[1]{\mathbf{#1}}
	\newcommand{\ib}[1]{\in\mathbb{#1}}
	\newcommand{\ic}[1]{\in\mathcal{#1}}
	\newcommand{\ca}[1]{\mathcal{#1}}
	\newcommand{\mrm}[1]{\mathrm{#1}}

	\title{Mechanical Power Modeling and Energy Efficiency Maximization for Movable Antenna Systems}
	\author{\IEEEauthorblockN{Xin Wei\IEEEauthorrefmark{1}, Weidong Mei\IEEEauthorrefmark{1}, Xuan Huang\IEEEauthorrefmark{2}, Zhi Chen\IEEEauthorrefmark{1}, and Boyu Ning\IEEEauthorrefmark{1}}
		\IEEEauthorblockA{\IEEEauthorrefmark{1}National Key Laboratory of Wireless Communications,}
		\IEEEauthorblockA{\IEEEauthorrefmark{2}School of Information and Communication Engineering,}
			University of Electronic Science and Technology of China, Chengdu, China.\\
		Email: xinwei@std.uestc.edu.cn, wmei@uestc.edu.cn, huang.xuan@std.uestc.edu.cn,\\ chenzhi@uestc.edu.cn, boydning@outlook.com}
	\maketitle
	\newcommand\blfootnote[1]{%
		\begingroup
		\renewcommand\thefootnote{}\footnote{#1}%
		\addtocounter{footnote}{-1}%
		\endgroup
	}
	\begin{abstract}
		Movable antennas (MAs) have recently garnered significant attention in wireless communications due to their capability to reshape wireless channels via local antenna movement within a confined region. However, to achieve accurate antenna movement, MA drivers introduce non-negligible mechanical power consumption, rendering energy efficiency (EE) optimization more critical compared to conventional fixed-position antenna (FPA) systems. To address this problem, we develop in this paper a fundamental power consumption model for stepper motor-driven MA systems by resorting to basic electric motor theory. Based on this model, we formulate an EE maximization problem by jointly optimizing an MA's position, moving speed, and transmit power. However, this problem is difficult to solve optimally due to the intricate relationship between the mechanical power consumption and the design variables. To tackle this issue, we first uncover a hidden monotonicity of the EE performance with respect to the MA's moving speed. Then, we apply the Dinkelbach algorithm to obtain the optimal transmit power in a semi-closed form for any given MA position, followed by an enumeration to determine the optimal MA position. Numerical results demonstrate that despite the additional mechanical power consumption, the MA system can outperform the conventional FPA system in terms of EE.
	\end{abstract}
	
	\begingroup
	\allowdisplaybreaks
	\section{Introduction}
	The movable antenna (MA) is an emerging technique that enables proactive channel reconfiguration by adjusting the positions of multiple antennas within a confined region~\cite{zhu2024movable,ning2024movable,zhu2025tutorial,zheng2025rotatable,shao2025tutorial}. In addition, MAs can alter the spatial correlations of steering vectors over different steering angles for more flexible beamforming. Owing to these advantages, MAs hold significant promise for a variety of wireless network scenarios, such as array signal processing  \cite{ma2024multi,wang2024flexible}, single- and multi-user multiple-input single-output (MISO) \cite{zhu2023movable,mei2024movable}, multiple-input multiple-output (MIMO) \cite{ma2024mimo}, cognitive radio \cite{wei2024joint}, wireless sensing~\cite{ma2024movable,wang2025antenna}, secure communications \cite{mei2024posistion}, among others. Moreover, integration of MAs and other emerging technologies, e.g., intelligent reflecting surface \cite{wei2025movable} and integrated sensing and communications (ISAC) \cite{lyu2025movable,qin2024cramer}, has also been explored in the literature. All of the above works demonstrate the significant performance gains of MAs over conventional fixed-position antennas (FPAs). However, despite these prior works, the energy efficiency (EE) of MA systems has not been thoroughly investigated. Particularly, to enable accurate position tuning and movement of MAs, their mechanical drivers (e.g., stepper motors) can introduce non-negligible additional power consumption. Hence, it remains unknown whether MAs can still outperform FPAs in terms of EE. Although some prior works (e.g., \cite{wu2024globally} and \cite{chen2025energy}) have investigated the EE of MA systems, they lack detailed power consumption modeling for the MA drivers, instead treating it as a constant regardless of the MAs’ movement speeds. Hence, it is essential to develop a more realistic and hardware-aware power consumption model for EE optimization in MA systems.
	
	\begin{figure}[!t]
		\centering
		\captionsetup{justification=raggedright,singlelinecheck=false}
		\centerline{\includegraphics[width=0.4\textwidth]{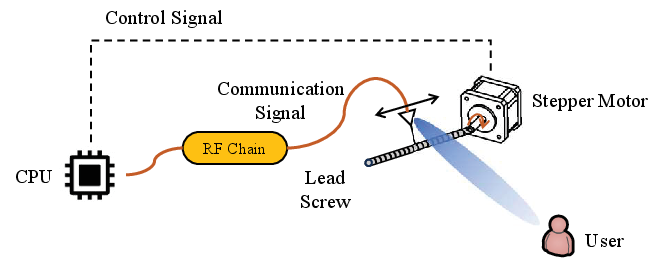}}
		\captionsetup{font=footnotesize}
		\vspace{-9pt}
		\caption{Stepper motor-driven MA system.}
		\label{Fig_SysModel}
		\vspace{-20pt}
	\end{figure}
	
	Motivated by the above, we study in this paper an EE maximization problem for a mechanically-driven single-MA system between an access point (AP) and a user, as shown in Fig.~\ref{Fig_SysModel}. The MA is driven by a stepper motor, which employs a lead screw as its linear actuator. To characterize the power consumption of the MA driver, we develop a fundamental power consumption model based on basic electric motor theory, where the power consumption of the stepper is calculated as the product of its angular speed and pull-out torque. Under this model, we formulate an EE maximization problem by jointly optimizing the MA's position, moving speed, and transmit power. However, this problem is challenging to be solved optimally due to the strong coupling among the design variables and highly nonlinear expression of the power consumption of the stepper motor. To address this, we first reveal a hidden monotonicity of the EE performance with respect to the MA's moving speed. Then, the Dinkelbach algorithm is applied to derive the optimal transmit power in a semi-closed form for any given MA position, followed by an enumeration to determine the optimal MA position. Numerical results demonstrate that the MA can significantly enhance the EE performance compared to the conventional FPA, despite the additional power consumption associated with MA movement.
	
	\section{System Model and Power Consumption Model}
	\subsection{System Model}
	As shown in Fig. 1, we consider an MA-aided communication system where an AP equipped with a single MA serves a user with a single FPA. The MA is mechanically driven via a stepper motor, which employs a lead screw with the outer radius of $l_0$ as its linear actuator, as illustrated in Fig. 1. We consider that the MA can adjust its position within a linear array with the length of $A$ in meter (m). Considering the discrete nature of stepper motors which rotate by a fixed step angle, the MA can only move in a discrete step size rather than continuously. Let $\omega_{D}$ denote the step angle of the stepper motor and the corresponding step size of the MA is expressed as $d_s=\omega_{D}l_0$. As a result, the continuous linear array is discretized into $J_x=\lfloor \frac{A}{d_s}\rfloor$ points as the MA's candidate positions, where $\lfloor\cdot\rfloor$ denotes the round down operation. Let $\ca{C}_t\triangleq\left\{0,d_s,2d_s,\cdots,(J_x-1)d_s\right\}$ denote the set of all candidate positions and $x_0$, $x_0\ic{C}_t$ denote the initial position of the MA.
	
	\begin{figure}[!t]
		\centering
		\captionsetup{justification=raggedright,singlelinecheck=false}
		\centerline{\includegraphics[width=0.4\textwidth]{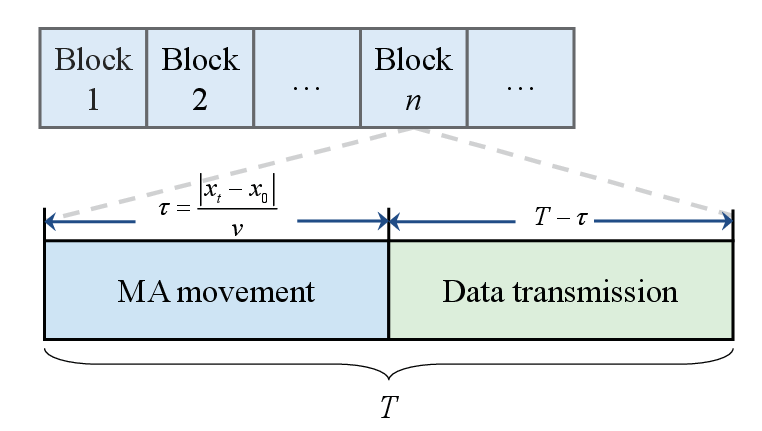}}
		\captionsetup{font=footnotesize}
		\vspace{-9pt}
		\caption{Two-stage transmission protocol in MA systems.}
		\label{Fig_Protocol}
		\vspace{-20pt}
	\end{figure}
	
	In this paper, we assume a block-fading channel model and focus on a given time block of duration $T$, which is divided into two stages, as illustrated in Fig.~\ref{Fig_Protocol}. Specifically, in the first stage, the stepper motor drives the MA to its destination (denoted as $x_t$, $x_t\ic{C}_t$ and to be optimized). We assume that the AP remains nearly inactive in this phase with negligible energy consumption; hence, the power consumption in this stage is mainly from the stepper motor. In addition, we assume a constant moving speed for the MA, denoted as $v$. To ensure that the MA can move to $x_t$ within the channel coherence time, its speed should satisfy $v\ge\frac{|x_t-x_0|}{T}$, and the time delay caused by antenna movement is given by $\tau=\frac{|x_t-x_0|}{v}$.\footnote{It is worth noting that for simplicity, we ignore the time duration and power consumption for accelerating and decelerating the MA at the beginning and end of the first stage, respectively, as they are much smaller than those during the constant-speed movement \cite{acarnley2002stepping}.} In the second stage, the data transmission is carried out, and the power consumption arises exclusively from the AP. \vspace{-8pt}
	
	\subsection{Power Consumption Model}
	According to the transmission protocol, the total power consumption of the system comprises the energy radiated for data transmission in the second stage and that consumed by the MA drivers in the first stage, i.e.,
	\begin{equation}\label{eqn_PwCsp}
		E_{\mrm{total}}\left(P,x_t,v\right)=\underset{\text{First stage}}{\underbrace{\tau P_M}}+\underset{\text{Second stage}}{\underbrace{\left(T-\tau\right)P_D}},\vspace{-5pt}
	\end{equation}
	where $P_M$ and $P_D$ denote the power consumption for the MA driver and data transmission, respectively. For the power consumption model during data transmission, it primarily includes the radiated power used for signal transmission and the static circuit power consumption, which is given by \cite{xu2013energy}
	\begin{equation}
		P_{D}=P+P_{s},
	\end{equation}
	where $P\le P_{\max}$ denotes the AP's transmit power with $P_{\max}$ denoting its maximum transmit power, and $P_{s}$ denotes the static circuit power consumption.
	
	Next, we develop the power consumption model of the MA driver. According to basic electric motor theory, the power consumption of the stepper motor primarily results from the mechanical work required to drive the load, which depends on the MA's moving speed $v$ and is given by \cite{acarnley2002stepping}
	\begin{equation}\label{eqn_MotorPower}
		P_M=\omega M(\omega)=\frac{v}{l_0}M\left(\frac{v}{l_0}\right),
	\end{equation}
	where $\omega=v/l_0$ denotes the angular velocity of the stepper motor. In addition, $M(\omega)$ represents the pull-out torque of the stepper motor and is characterized as \cite{acarnley2002stepping}
	\begin{equation}\label{eqn_Torque}
		M(\omega)=\frac{p\psi_{\mrm{M}}V}{\sqrt{R^2+\omega^2L^2}}-\frac{p\omega\psi_{\mrm{M}}^2R}{R^2+\omega^2L^2},
	\end{equation}
	where $p$ denotes the number of the rotor teeth, $\psi_{\mrm{M}}$ denotes the peak magnet flux linking each winding, $V$ denotes the voltage, $R$ denotes the phase resistance, and $L$ denotes the phase induction.
	
	\begin{figure}[!t]
		\centering
		\subfloat[]{
			\includegraphics[width=0.5\linewidth]{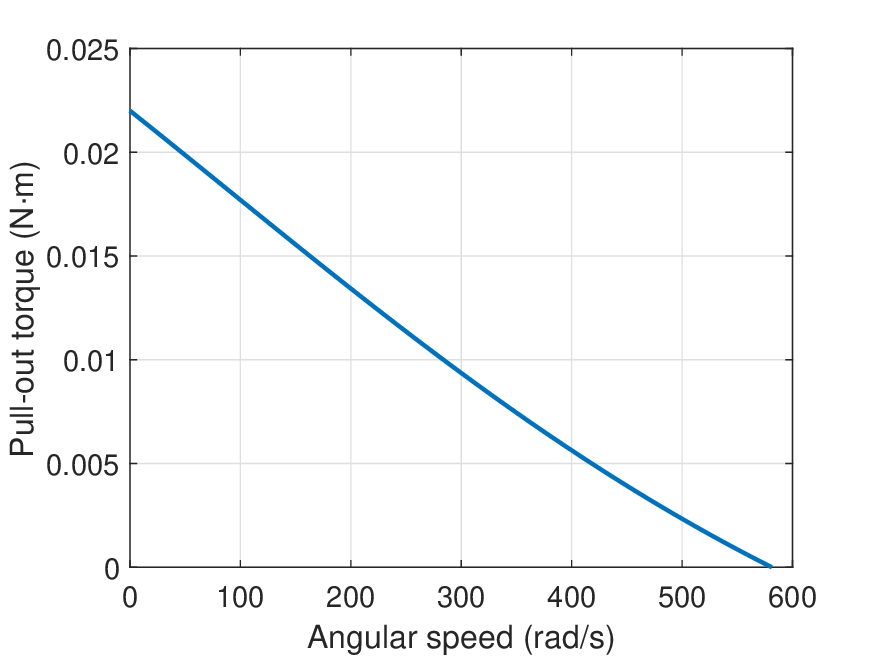}
		}
		\hfil
		\hspace{-20pt}
		\subfloat[]{
			\includegraphics[width=0.5\linewidth]{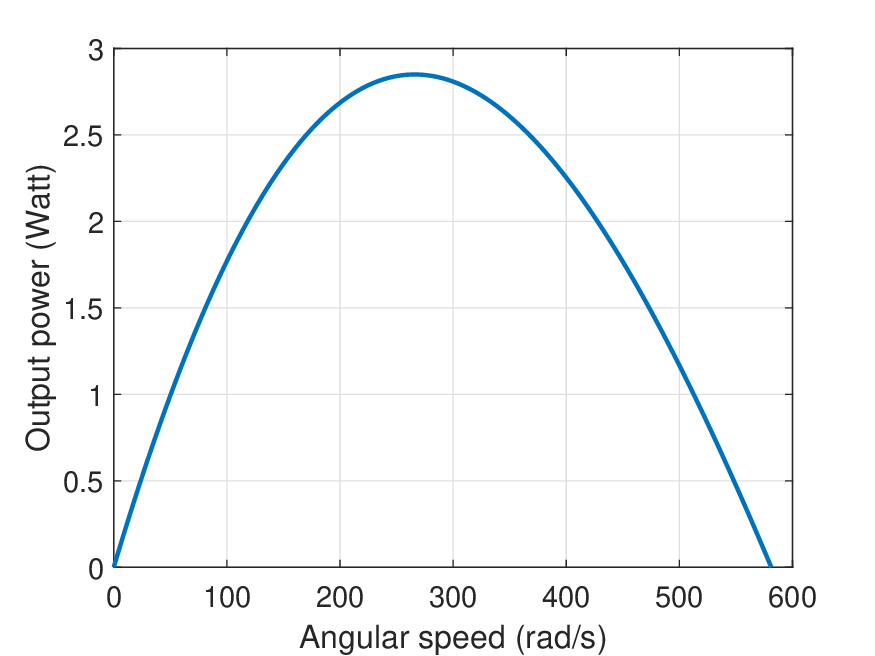}
		}
		\captionsetup{font=footnotesize}
		\caption{(a) Pull-out torque; (b) Output power versus the angular speed of the stepper motor.}
		\vspace{-20pt}
		\label{Fig_Motor}
	\end{figure}
	
	To facilitate a deeper understanding of the mechanical power consumption, we plot in Figs.~\ref{Fig_Motor}(a) and \ref{Fig_Motor}(b) the pull-out torque in \eqref{eqn_Torque} and the power consumption in \eqref{eqn_MotorPower} versus the speed of its driven load, i.e., $v$, respectively. The voltage is $V=11.94$ volt (V), the phase resistance is $R=75$ Ohm ($\Omega$), the phase inductance is $L=65.6$ millihenry (mH), the number of rotor teeth is $p=6$, the peak magnet flux linking each winding is $\psi_{M}=0.023$ Weber (Wb), and the radius of the lead screw is $l_0=5$ millimeter (mm). It can be observed from Fig.~\ref{Fig_Motor}(a) that the pull-out torque of the stepper motor gradually decreases with its angular velocity $\omega$. This is because the pull-out torque produced by a stepper motor is proportional to the current through the coils. As $\omega$ increases, the back electromotive force (EMF) generated by the motor's windings increases, which acts against the applied voltage and reduces the effective voltage driving the current through the motor coils. On the other hand, stepper motors are inductive loads. The inductive resistance of the coils also increases with $\omega$, which limits the current that can be supplied to the windings. Hence, it is shown in Fig.~\ref{Fig_Motor}(b) that the power consumption of a stepper motor first increases with $v$, achieving its maximum value, and then decreases, rather than remaining constant. It is also worth noting that the power consumption drops to zero when the angular speed reaches its maximum value (denoted by $\omega_{M}$), corresponding to the no-load condition, which is typically unachievable in practice due to the non-negligible weight of the MA. Therefore, we set a maximum achievable angular speed $\omega_{\max}$, $\omega_{\max}<\omega_{M}$ for the stepper motor in this paper, accounting for the load of the MA. As such, the maximum moving speed of the MA is given by $v_{\max}=\omega_{\max}l_0$.\vspace{-8pt}
	
	\section{Problem Formulation}
	\vspace{-8pt}Based on \eqref{eqn_PwCsp}-\eqref{eqn_Torque}, we formulate the EE maximization problem for the considered MA system in this section. Let $h(x_t)\ib{C}$ denote the AP-user channel with respect to the position $x_t$, $x_t\ic{C}_t$. Then, the achievable rate of the user is given by
	\begin{equation}
		R\left(x_t,P\right)=\log_2\left(1+\frac{P}{\sigma^2}\left|h(x_t)\right|^2\right),
	\end{equation}
	where $\sigma^2$ denotes the average noise power at the user. As such, the energy efficiency of the considered single-MA system within a channel coherence block can be expressed as
	\begin{align}
		\mrm{EE}(P,x_t,v)&=\frac{(T-\tau)R\left(x_t,P\right)}{E_{\mrm{total}}\left(P,x_t,v\right)}\label{eqn_EE}\\
		&=\frac{\left(vT-|x_t-x_0|\right)R\left(x_t,P\right)}{(vT-|x_t-x_0|)(P+P_s)+P_M|x_t-x_0|}.\nonumber
	\end{align}
	It is observed from \eqref{eqn_EE} that the EE performance depends on the speed of the MA, $v$. As $v$ increases, the movement delay $\tau$ will decrease, which helps increase the achievable rate in the numerator of \eqref{eqn_EE}. However, it remains unclear whether the total power consumption in the denominator of \eqref{eqn_EE} will increase or not. As such, there exists a non-trivial relationship between the EE in \eqref{eqn_EE} and the MA's speed $v$. Furthermore, the EE in \eqref{eqn_EE} also depends on the destination position $x_t$, as it affects both the communication channel power gain and the movement delay for a given $v$. Last but not least, the AP's transmit power $P$ can also affect \eqref{eqn_EE} due to the different scaling rate of its denominator versus its numerator.
	
	Hence, in this paper, we aim to maximize \eqref{eqn_EE} by jointly optimizing the destination position of the MA $x_t$, the AP's transmit power $P$, and the moving speed of the MA $v$. The associated optimization problem can be formulated as
	\begin{subequations}
		\begin{align}
			(\mrm{P}1)\quad\underset{P,x_t,v}{\max}\quad&\mrm{EE}(P,x_t,v)\nonumber\\
			\mrm{s.t.}\quad &0\le P \le P_{\max},	\label{eqn_C1}\\
			&\frac{|x_t-x_0|}{T}\le v \le v_{\max},	\label{eqn_C2}\\
			& x_t\ic{C}_t. \label{eqn_C3}
		\end{align}
	\end{subequations}
	Note that for (P1), if the optimized destination antenna position $x_t$ is set identical to the initial position $x_0$, the MA system reduces to the conventional FPA system. As such, the optimal EE value of (P1) is ensured to be no worse than that by an FPA system. Moreover, if the channel coherence time $T$ is sufficiently large, e.g., $T \rightarrow \infty$, the EE in \eqref{eqn_EE} will degrade to conventional EE without accounting for the mechanical power consumption, i.e.,
	\begin{equation}\label{eqn_EE_infty}
		\mrm{EE}(P,x_t,v)\approx\frac{R(x_t,P)}{P+P_s},
	\end{equation}
	as studied in \cite{wu2024globally} and \cite{chen2025energy}. It follows that a different MA position from that in \cite{wu2024globally} and \cite{chen2025energy} is generally needed considering the mechanical power consumption, especially if the channel coherence time is not long.
	
	Note that to characterize the fundamental limit of the energy efficiency of the considered MA system, we assume that all required channel state information is available via the existing channel estimation techniques for MA systems; see \cite{ma2023compressed} and \cite{zhang2024channel}. However, (P1) remains difficult to be solved optimally due to the fractional form of the objective function, and the nonlinear and intricate expression of $P_M$. To address these difficulties, we propose an efficient solution to (P1) in the next section.
	
	\section{Proposed Solution to (P1)}
	\subsection{Optimal MA Moving Speed}
	First, to cope with the highly nonlinear expression of $P_M$ with respect to $v$, we introduce the following proposition to capture the relationship between EE and MA moving speed.
	\begin{proposition}
		For any given MA's destination $x_t$ and the AP's transmit power $P$, the EE in \eqref{eqn_EE} monotonically increases with $v$.
	\end{proposition}
	\begin{IEEEproof}
		First, we recast the EE in \eqref{eqn_EE} as a more tractable form:
		\begin{equation}\label{eqn_EE_2}
			\begin{aligned}
				&\mrm{EE}(v)\\
				&=\frac{\left(vT-|x_t-x_0|\right)R\left(x_t,P\right)}{(vT-|x_t-x_0|)(P+P_s)+P_M|x_t-x_0|}\\
				&=\frac{R\left(x_t,P\right)}{P+P_s+f(v)|x_t-x_0|},
			\end{aligned}
		\end{equation}
		where $f(v)\triangleq\frac{P_M}{vT-|x_t-x_0|}$. It is evident that EE increases with $v$ if $f(v)$ decreases with $v$. The first-order derivative of $f(v)$ is given by
		\begin{equation}
			\frac{\partial f(v)}{\partial v}=\frac{v\frac{\partial M(\omega)}{\partial \omega}\left(vT-|x_t-x_0|\right)-l_0M\left(\frac{v}{l_0}\right)|x_t-x_0|}{l_0^2\left(vT-|x_t-x_0|\right)^2}.
		\end{equation}
		Note that $vT-|x_t-x_0|>0$ due to the constraints in \eqref{eqn_C2}. Moreover, for practical stepper motors, its pull-out torque decreases with its angular velocity, as depicted in Fig. 2(a). Hence, we have $\frac{\partial M(\omega)}{\partial \omega}<0$. Therefore, we can conclude that $\frac{\partial f(v)}{\partial v}<0$; thus, $f(v)$ monotonically decreases with $v$. This completes the proof.
	\end{IEEEproof}
	
	Proposition 1 indicates that to maximize the EE of the considered mechanically-driven MA system, the stepper motor should always operate at its maximum speed, i.e., $v=v_{\max}$, which also helps prolong the time for data transmission. By substituting $v=v_{\max}$ into (P1), (P1) can be simplified as
	\begin{subequations}
		\begin{align}
			(\mrm{P}2)\quad\underset{P,x_t}{\max}\quad&\mrm{EE}(P,x_t,v_{\max})\nonumber\\
			\mrm{s.t.}\quad &\frac{|x_t-x_0|}{T}\le v_{\max},\label{eqn_C4}\\
			&\eqref{eqn_C1},\eqref{eqn_C3}.\nonumber
		\end{align}
	\end{subequations}
	
	\subsection{Proposed Algorithm to (P2)}
	Next, we show that for any given $x_t$, we can derive the optimal $P$ for (P2) in a semi-closed form via the Dinkelbach algorithm. Specifically, if $x_t$ is fixed, (P2) reduces to the following optimization problem:
	\begin{subequations}
		\begin{align}
			(\mrm{P}3)\,\,\underset{P}{\max}\,\,&\frac{\left(v_{\max}T-|x_t-x_0|\right)R\left(x_t,P\right)}{(v_{\max}T-|x_t-x_0|)(P+P_s)+G(x_t)},\,\,\mrm{s.t.}\,\,\eqref{eqn_C1},\nonumber
		\end{align}
	\end{subequations}
	where $G(x_t)\triangleq P_{M}(v_{\max})|x_t-x_0|$. Note that (P3) is a classical fractional programming (FP) problem, for which the Dinkelbach's algorithm can be employed. Specifically, in the $l$-th iteration of the Dinkelbach's algorithm, (P3) is transformed into the following subtractive form, i.e.,
	\begin{equation}\label{eqn_func_P3-1}
			(\mrm{P}3-l)\quad\underset{P}{\max}\quad f^{(l)}(P)\quad\mrm{s.t.}\quad\eqref{eqn_C1},\nonumber
	\end{equation}
	where
	\begin{align}
		f^{(l)}(P)&\triangleq\left(v_{\max}T-|x_t-x_0|\right)R(x_t,P)-\eta^{(l-1)}\times\\
		&\quad\,\left[(v_{\max}T-|x_t-x_0|)(P+P_s)+G(x_t)\right]\nonumber
	\end{align}
	and
	\begin{equation}\label{eqn_Eta}
		\eta^{(l-1)}\triangleq\frac{\left(v_{\max}T-|x_t-x_0|\right)R(x_t,P^{(l-1)})}{(v_{\max}T-|x_t-x_0|)(P^{(l-1)}+P_s)+G(x_t)}
	\end{equation}
	denotes the EE value obtained in the $(l-1)$-th iteration with $P^{(l-1)}$ denoting the optimal transmit power obtained in this iteration. Note that $f^{(l)}(P)$ is a concave function in $P$. By setting $\frac{\partial f^{(l)}(P)}{\partial P}=0$, the optimal transmit power that maximizes \eqref{eqn_func_P3-1} can be obtained as
	\begin{equation}\label{eqn_Opt_TxPw}
		P^{(l)}=\min\left(\left[\frac{1}{\eta^{(l-1)}\ln2}-\frac{\sigma^2}{|h(x_t)|^2}\right]^+,P_{\max}\right),
	\end{equation}
	where $\left[z\right]^+=\max(z,0)$. Based on \eqref{eqn_Opt_TxPw} and \eqref{eqn_Eta}, we can compute the value of $\eta^{(l)}$ and proceed to the $(l+1)$-th Dinkelbach iteration. It can be shown that $\eta^{(l)}$ will monotonically increase with the iteration number and converge to an optimal solution to (P3) \cite{xu2013energy}. 
	
	Denote by $P(x_t)$ the optimal transmit power for (P3) by the Dinkelbach algorithm for any given $x_t$. Then, we can perform an enumeration of all candidate positions in $\ca{C}_t$ to obtain the optimal MA position for (P2) as
	\begin{equation}
		x_t^*=\arg\underset{x_t\ic{C}_t}{\max}\,\,\mrm{EE}\left(P(x_t),x_t,v_{\max}\right).
	\end{equation}
	The overall procedures of the proposed algorithm to solve (P2) are summarized in Algorithm 1. It can be shown that the computational complexity of Algorithm 1 is given by $\ca{O}(J_xI_1)$, where $I_1$ denotes the number of iterations of the Dinkelbach algorithm for solving (P3).
	
	\begin{algorithm}[!t]
		\caption{Proposed Algorithm for Solving (P2)}
		\label{alg_SU}
		\begin{algorithmic}[1]
			\STATE Initialize the step size $d_s$, the number of all candidate points $J_x=\lfloor\frac{A}{d_s}\rfloor$, and the convergence accuracy $\epsilon$.
			\STATE Initialize $\mrm{EE}^{\star}=0$, $x_t^{\star}=0$ and $P^{\star}=0$.
			\FOR{$j=1\rightarrow J_x$}
			\STATE Set $x_t=(j-1)d_s$, $l=1$, and $P^{(l-1)}=P_{\max}$.
			\IF{$|x_t-x_0|/v_{\max}<T$}
			\STATE $\mrm{EE}^{\star}=-\infty$.
			\ENDIF
			\STATE Initialize $\eta^{(l-1)}$ according to \eqref{eqn_Eta}.
			\REPEAT
			\STATE Set $l=l+1$.
			\STATE Calculate $P^{(l)}$ according to \eqref{eqn_Opt_TxPw}.
			\STATE Calculate $\eta^{(l)}$ according to \eqref{eqn_Eta}.
			\UNTIL{$|\eta^{(l)}-\eta^{(l-1)}|<\epsilon$.}
			\IF{$\eta^{(l)} > \mrm{EE^{\star}}$}
			\STATE Update $\mrm{EE^{\star}}=\eta^{(l)}$, $x_t^{\star}=x_t$, and $P^{\star}=P^{(l)}$.
			\ENDIF
			\ENDFOR
			\STATE Output $x_t^{\star}$ and $P^{\star}$ as the optimized solutions to (P2).
		\end{algorithmic}
	\end{algorithm}
	
	\section{Numerical Results}
	In this section, we present numerical results to validate the efficacy of our proposed algorithm. Unless otherwise specified, the simulation parameters are set as follows. The carrier wavelength is $\lambda=0.06$ meter (m). The length of the linear array is $A=2\lambda=0.12$ m. The initial position of the MA is $x_0=A/2$. The AP's static circuit power consumption is $P_s=30$ dBm, and its maximum transmit power is $P_{\max}=46$ dBm. The time duration of a channel coherence block is $T=50$ ms. We consider the field-response-based channel model presented in \cite{zhu2025tutorial} for MAs in the simulation. The distance from the AP to the user is $d$ = 30 m, and the path loss exponent for the AP-user channel is $\alpha=2.8$. The number of transmit paths in the AP-user channel is $L=4$, and their path gains are assumed to follow the circle symmetric complex Gaussian (CSCG) distribution, i.e., $g_k\sim\ca{CN}(0,\rho d^{-\alpha}/L)$, $k=1,2,\cdots,L$, where $\rho$ represents the path loss at the reference distance of 1 m. The angles of departure (AoDs) for these paths are assumed to be independent and identically distributed variables following the uniform distribution within $[-\frac{\pi}{2},\frac{\pi}{2}]$. The average noise power is $\sigma^2=-80$ dBm. Moreover, we consider the AM2224 high-speed stepper motor in this simulation \cite{am2224}, with the same parameters as those adopted to Fig. 3. The maximum angular speed of the stepper motor is $\omega_{\max}=552$ rad/s. The step angle and radius of the lead screw is $\omega_{D}=\frac{\pi}{12}$ rad/s and $l_0=5$ mm, respectively. Hence, the step size of the stepper motor is $d_s=\omega_{D}l_0\approx1.2$ mm. All the results are averaged over 1000 independent channel realizations.
	
	Furthermore, we consider the following benchmark schemes for performance comparison:
	\begin{enumerate}
		\item \textbf{Benchmark 1: Achievable rate maximization.} In this benchmark, the antenna position $x_t$ is optimized to solely maximize the achievable rate $R(x_t)$. Thus, the AP's transmit power is fixed as $P=P_{\max}$.
		\item \textbf{Benchmark 2: Fixed motor power consumption.} In this benchmark, the output power of the stepper motor is fixed as a constant denoted by $P_M=\frac{v_{\max}}{2l_0}M\left(\frac{v_{\max}}{2l_0}\right)$, and the antenna position $x_t$ and the AP's transmit power $P$ are optimized via Algorithm 1.
		\item \textbf{Benchmark 3: FPA.} In this benchmark, the antenna position is fixed as $x_t=x_0$, and the AP's transmit power is optimized via the Dinkelbach's algorithm.
	\end{enumerate}
	
	\begin{figure}[t]
		\centering
		\captionsetup{justification=raggedright,singlelinecheck=false}
		\centerline{\includegraphics[width=0.4\textwidth]{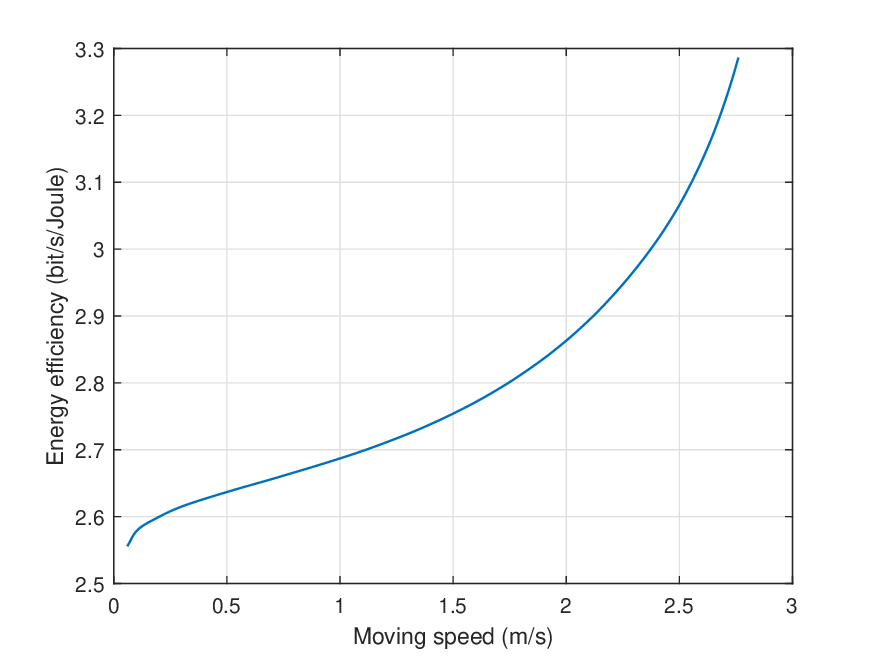}}
		\captionsetup{font=footnotesize}
		\caption{EE performance of the proposed algorithm versus the MA's moving speed.}
		\label{Fig_Velocity}
		\vspace{-15pt}
	\end{figure}
	First, to validate Proposition 1, we show in Fig.~\ref{Fig_Velocity} the EE performance of the proposed algorithm versus the MA's moving speed $v$. It can be observed that the EE performance monotonically increases with $v$, which is consistent with our analysis in Proposition 1.
	
	\begin{figure}[t]
		\centering
		\captionsetup{justification=raggedright,singlelinecheck=false}
		\centerline{\includegraphics[width=0.4\textwidth]{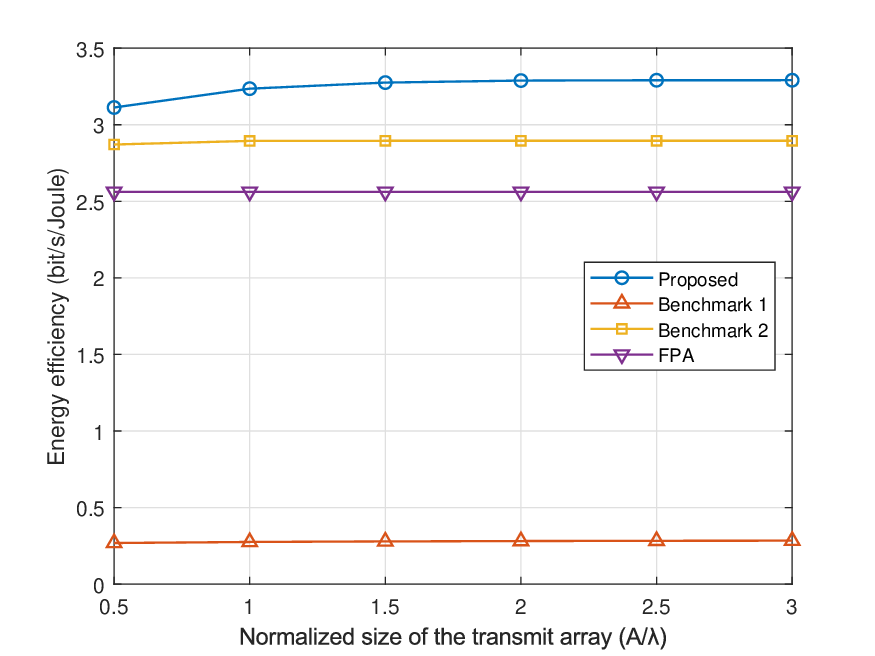}}
		\captionsetup{font=footnotesize}
		\caption{EE performance versus the normalized size of the transmit array.}
		\label{Fig_TxRegion}
		\vspace{-15pt}
	\end{figure}
	Next, we plot in Fig.~\ref{Fig_TxRegion} the EE performance versus the normalized size of the transmit array (i.e., $A/\lambda$). It is observed that the EE performance of the proposed algorithm increases with the array size and outperforms the other three benchmarks. Nonetheless, its EE performance saturates as $A\ge2\lambda$, suggesting that a finite movable region suffices achieve the optimal EE performance. Furthermore, it is observed that the performance of the three benchmark schemes remains nearly constant as the array size varies. This is because they are optimized to always transmit at the maximum power rendering the EE performance dominated by the maximum transmit power $P_{\max}$.
	
	\begin{figure}[t]
		\centering
		\captionsetup{justification=raggedright,singlelinecheck=false}
		\centerline{\includegraphics[width=0.4\textwidth]{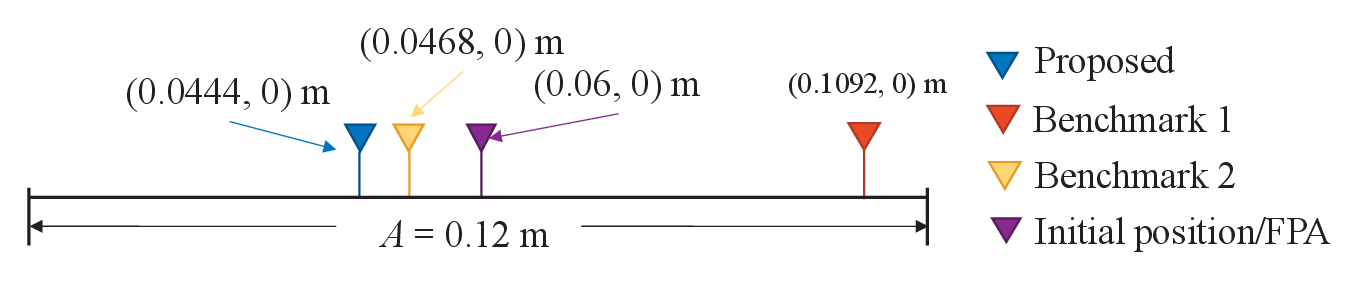}}
		\captionsetup{font=footnotesize}
		\caption{The optimized positions by all schemes.}
		\label{Fig_OptPos}
		\vspace{-15pt}
	\end{figure}
	
	To gain more insights, we plot in Fig.~\ref{Fig_OptPos} the optimized MA destination positions by all schemes, marked by different colors. It is observed that the optimized positions by the proposed algorithm and Benchmark 2 are located closely to the initial position, thereby reducing the energy consumption and movement delay due to the long-distance movement. In contrast, the optimized position in Benchmark 1 is farthest from the initial position among all considered schemes, so as to achieve the best channel condition. This, however, results in the lowest EE performance as observed in Fig.~\ref{Fig_TxRegion} due to the high mechanical power consumption.
	
	\begin{figure}[t]
		\centering
		\captionsetup{justification=raggedright,singlelinecheck=false}
		\centerline{\includegraphics[width=0.4\textwidth]{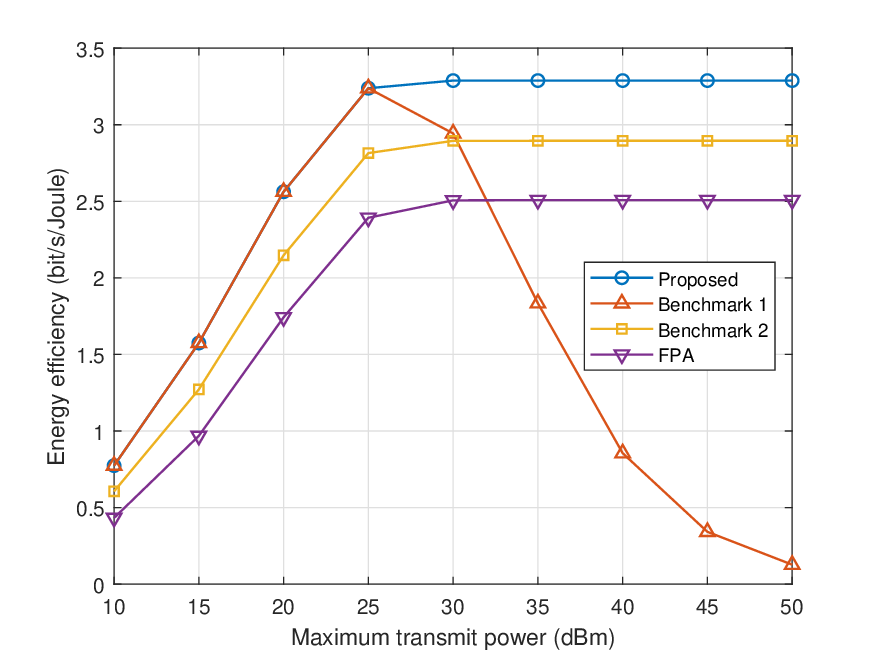}}
		\captionsetup{font=footnotesize}
		\caption{EE performance versus the AP's maximum transmit power.}
		\label{Fig_TxPower}
		\vspace{-20pt}
	\end{figure}
	Fig.~\ref{Fig_TxPower} shows the EE performance versus the AP's maximum transmit power $P_{\max}$. It is observed that the EE performance of all schemes increases with $P_{\max}$ when $P_{\max}\le25$ dBm. This is because the EE performance is mainly dominated by the achievable rate in the low transmit power regime. Hence, all schemes should transmit at the maximum power to maximize the achievable rate. However, when $P_{\max} > 25$ dBm, the total power consumption plays a more significant role. As a result, transmitting at the maximum power may result in a significant loss in EE, as observed from the performance of Benchmark 1. In contrast, the performance of the other three schemes is observed to remain constant as $P_{\max}$ increases, suggesting that their transmit powers are unchanged with $P_{\max}$ to maximize EE.
	
	\begin{figure}[t]
		\centering
		\captionsetup{justification=raggedright,singlelinecheck=false}
		\centerline{\includegraphics[width=0.4\textwidth]{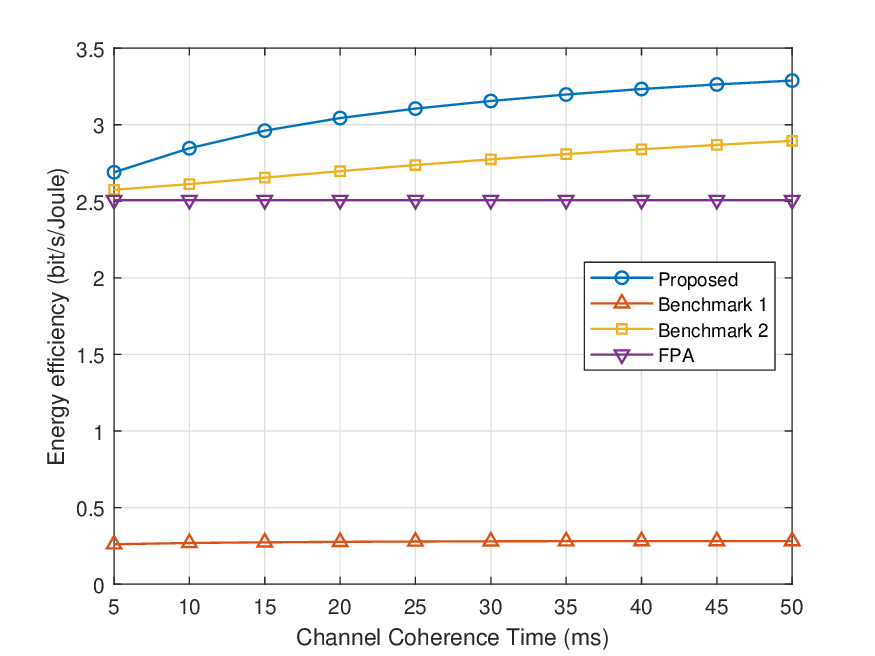}}
		\captionsetup{font=footnotesize}
		\caption{EE performance versus the channel coherence time.}
		\label{Fig_TotalTime}
		\vspace{-15pt}
	\end{figure}
	
	Fig.~\ref{Fig_TotalTime} depicts the EE performance versus the channel coherence time $T$. It is observed that the EE performance of all schemes (except the FPA benchmark) improves with increasing $T$, as this increases the MA's maximum movable distance as seen from \eqref{eqn_C4}. In contrast, the EE performance of the FPA benchmark keeps constant as $T$ increases. In addition, it is observed that the increase in the EE performance by the proposed scheme decreases with increasing $T$ and ultimately converges. This is because as $T$ is sufficiently large, the EE in \eqref{eqn_EE} will degrade to \eqref{eqn_EE_infty}, which is regardless of $T$, as discussed at the end of Section III.
	
	\begin{figure}[t]
		\centering
		\captionsetup{justification=raggedright,singlelinecheck=false}
		\centerline{\includegraphics[width=0.4\textwidth]{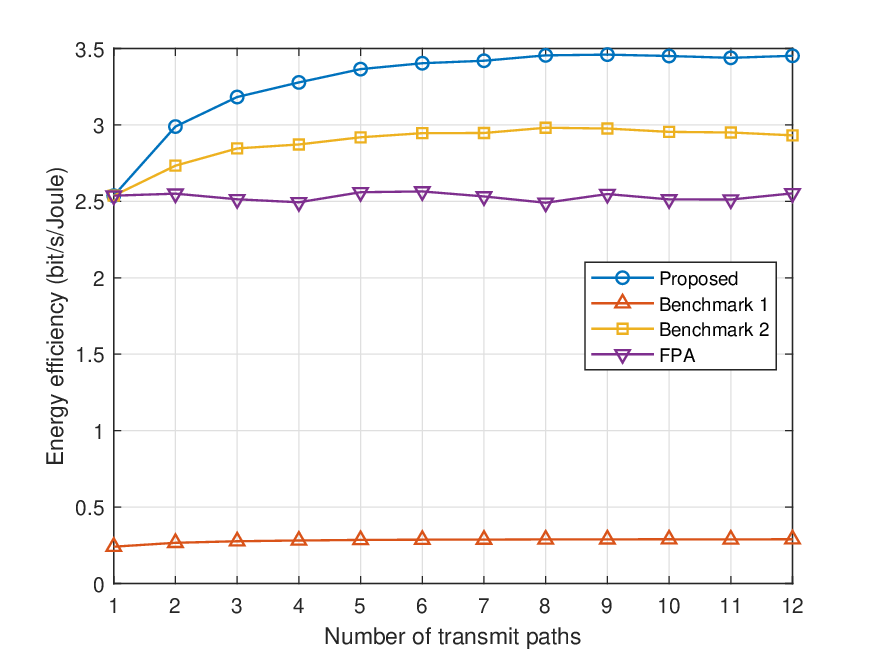}}
		\captionsetup{font=footnotesize}
		\caption{EE performance versus the number of transmit paths.}
		\label{Fig_PathNums}
		\vspace{-15pt}
	\end{figure}
	
	Lastly, Fig.~\ref{Fig_PathNums} shows the EE performance versus the number of transmit paths in the AP-user channel (i.e., $L$). It is observed that the EE performance of all schemes (except the FPA benchmark) improves as the number of paths increases. This is because when $L$ increases, the channel power gain within the movable region experiences more significant fluctuation, which yields a more significant spatial diversity gain. This helps shorten the movement time to identify the optimal antenna position that balances rate performance and energy consumption. In addition, the EE performance of the proposed algorithm, Benchmark 1 and the FPA benchmark is observed to be identical for $L=1$, due to the constant channel power gain within the movement region in this case.
	
	\section{Conclusions}
	In this paper, we investigated an EE maximization problem for a stepper motor-driven MA system. First, we developed a mechanical power model for the considered system. Based on this model, we revealed that the maximum EE performance can be achieved at the maximum moving speed of the MA. The optimal transmit power and MA destination position were then obtained by combining the Dinkelbach's algorithm and a full enumeration. Numerical results demonstrated that maximizing rate performance only may lead to a significant loss in EE. However, by properly optimizing the antenna position to balance rate performance and total energy consumption, the MA system can still outperform the FPA system.
	
	\bibliography{MA_EE.bib}
	\bibliographystyle{IEEEtran}
\end{document}